\documentclass[conference,a4paper]{APSIPA2021}
\usepackage{multirow}
\usepackage{graphicx}
\usepackage{amsmath}
\usepackage[psamsfonts]{amssymb}
\usepackage{amsxtra}
\usepackage{threeparttable}
\usepackage{makecell, multirow, lineno, hyperref}

\usepackage{geometry}
\usepackage{fancyhdr}
\renewcommand{\headrulewidth}{0pt}
\renewcommand{\footrulewidth}{0pt}
\fancypagestyle{firststyle} {
    \fancyhf{}
    \fancyhead[L]{Proceedings of 2022 APSIPA Annual Summit and Conference}
    \fancyhead[R]{7-10 November 2022, Chiang Mai, Thailand}
    
    \fancyfoot[L]{978-616-590-477-3 ©2022 APSIPA}

    \fancyfoot[R]{APSIPA ASC 2022}
}
\fancypagestyle{fancy} {
    \fancyhf{}
    \fancyhead[L]{Proceedings of 2022 APSIPA Annual Summit and Conference}
    \fancyhead[R]{7-10 November 2022, Chiang Mai, Thailand}
    
}

\begin{document}

\title{Speech Enhancement with Perceptually-motivated Optimization and Dual Transformations}

\author{%
\authorblockN{%
Xucheng Wan, Kai Liu, Ziqing Du, Huan Zhou
}
\authorblockA{%
Artificial Intelligence Application Research Center, Huawei Technologies \\Shenzhen, PRC\\
E-mail: {\{wanxucheng, liukai89, duziqing1, zhou.huan\}@huawei.com}}
}

\maketitle

\begin{abstract}
To address the monaural speech enhancement problem, numerous research studies have been conducted to enhance speech via operations either in time-domain on the inner-domain learned from the speech mixture or in time--frequency domain on the fixed full-band short time Fourier transform (STFT) spectrograms. Very recently, a few studies on sub-band based speech enhancement have been proposed. By enhancing speech via operations on sub-band spectrograms, those studies demonstrated competitive performances on the benchmark dataset of DNS2020. Despite attractive, this new research direction has not been fully explored and there is still room for improvement. As such, in this study, we delve into the latest research direction and propose a sub-band based speech enhancement system with perceptually-motivated optimization and dual transformations,  called PT-FSE. Specially, our proposed PT-FSE model improves its backbone, a full-band and sub-band fusion model, by three efforts. First, we design a frequency transformation module that aims to strengthen the global frequency correlation. Then a temporal transformation is introduced to capture long range temporal contexts. Lastly, a novel loss, with leverage of properties of human auditory perception, is proposed to facilitate the model to focus on low frequency enhancement. To validate the effectiveness of our proposed model, extensive experiments are conducted on the DNS2020 dataset. Experimental results show that our PT-FSE system achieves substantial improvements over its backbone, but also outperforms the current state-of-the-art while being 27\% smaller than the SOTA. With average NB-PESQ of 3.57 on the benchmark dataset, our system offers the best speech enhancement results reported till date.
\end{abstract}

 
\section{Introduction}

In many speech related applications, such as mobile communication and human-machine interfaces, the received speech often contains background noise interference, which degrades perceptual speech quality and intelligibility. To address the problem, speech enhancement (SE) system is often expected with objective of reconstructing the target clean speech from the noisy recording. Despite the growing interests, SE remains a challenging task, especially when the noise degraded mixture is recorded in single-channel with low signal-to-noise ratios (SNRs).

Over the past decade, numerous deep learning (DL)-based SE networks have been explored in the literature. With abundant paired data from clean and noisy speech, a DL-based SE neural network can be trained in a supervised fashion and achieve decent performance. 

Based on their operation domain, most DL-based SE networks can be roughly classified into two branches, which are time domain enhancement and time--frequency domain enhancement, with their respective pros and cons. 
One branch of these works operates in time domain \cite{Luo2017, zhang20m_interspeech,Luo2019, Luo20, zhang21}. They can either directly learn the regression function from the mixture to rebuild the target signal, or estimate target mask in the transformed inner-domain learned from the mixture. One of the most well-known works is time-domain audio separation network (TasNet) \cite{Luo2017}. It uses a convolutional encoder-decoder framework as an substitute of time--frequency (TF) transformation and perform source separation by estimating a weighting function for the encoder output at each time step. As a variant of TasNet, the same research group later proposed dual-path RNN (DPRNN) \cite{Luo20} for a good trade-off between computational requirements and performance. 

As a different strategy, the other branch of SE studies conducts enhancement in TF domain using spectrogram \cite{Hummersone14, tan18,choi19, Tan19,hu20g,Lv21, Hao21, Li19}. Using two-dimensional spectrogram representation of the mixture, the target signal, theoretically, could be faithfully recovered if both magnitude and phase are enhanced jointly. An exemplary technology is PHASEN \cite{Yin20}, which uses two parallel streams (one for amplitude mask, the other for phase prediction) and introduces information exchange between them for better phase estimation.

Among some recently proposed TF-based SE studies, two interesting and attractive research trends have emerged. Some studies propose the complex-valued model and achieve competitive performance, such as deep complex convolution recurrent network (DCCRN) \cite{hu20g}. By adopting a complex network with a complex target, it models the correlation between magnitude and phase via the simulation of complex multiplication, and ranks first in MOS evaluation in the DNS 2020 challenge \cite{dns2020}. Its follow-up research, called DCCRN+ \cite{Lv21}, further boosts performance of the DCCRN by introducing revisions on learnable sub-band split and complex TF-LSTM. 

In contrast to most majority TF-based methods that perform full-band spectrum enhancement, very recently, another line of research has emerged to split spectrogram of the mixture into multiple frequency bands and enhance each subband individually. Although such an idea has not been fully explored, a few very recent works on this direction \cite{Li19, Hao21, fullsubnet+} achieve unprecedented performances on the benchmark DNS dataset. 

Inspired by the pioneering subband-based SE studies, in this paper, we attempt to investigate the promising but underexplored topic. Firstly, we think the idea to enhance each subband individually sounds appealing, as it naturally in line with the fact that practical noise is mostly colored which affects speech spectrum differently at various frequencies. On the other hand, although performance gains are reported, we believe that there are limitations and still have room for improvement. In particular, we argue that the original FullSubNet model \cite{Hao21} may suffer two issues, model sensitivity among different frequency ranges is not fully developed and loss from each frequency subband may not yield equal contribution to the system, which are further elaborated in the following sections.

Motivated by the observations, in this study, building upon the baseline architecture of FullSubNet, we design a new SE network, Perceptive-motivated Optimization and Dual-Transformation basd Full-Sub band Speech Enhancement (PT-FSE), to address these issues, which improves the baseline with leverage of the TF contextual information and the prior knowledge of human perception.  

In all, in this study, we made the following contributions: i) propose to strengthen the spectrogram features with dual-transformation structures ; ii) design a novel loss to explicitly optimize for these important acoustic properties and force the network to pay attention to these low-level features; and iii) our proposed PT-FSE surpasses the baseline system but also outperforms the current state-of-the-art on the DNS challenge dataset.

The rest of the paper is organized as follows. Research works related to subband-based SE are introduced in Section 2. Section 3 describes our proposed system. Experimental results are
reported and analyzed in Section 4. Finally We conclude the paper in Section 5.
\section{Related Works}
In this section, we briefly introduce as two exemplary models related to the subband-based SE. 

\subsection{FullSubNet}
\begin{figure}[t]
  \centering
  \includegraphics[scale=0.4]{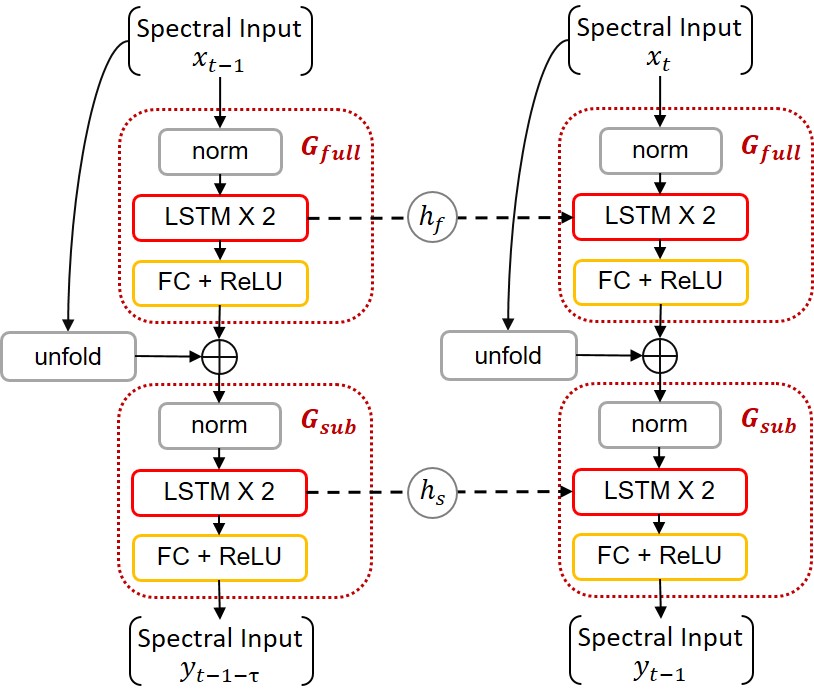}
  \caption{Structure of FullSubNet}
  \label{fig:FSN}
\end{figure}

Motivated to integrate complementary fullband and subband information for SE, FullSubNet \cite{Hao21} is proposed as a fullband and subband fusion model. Its core network architecture, as illustrated in fig. \ref{fig:FSN}, is composed of sequentially connected two modules, full-band network $G_{\mathrm{full}}$ and sub-band network $G_{\mathrm{sub}}$. The network takes STFT magnitude spectrogram sequence $X \in \mathcal{R}^{F*T}$ as input (where $F$ and $T$ denote the number of frequency bins and frames respectively), and predicts complex mask sequence $y\in \mathcal{R}^{2*F*T}$ with latency of $\tau$ time steps (here $\tau$ denotes a small latency to meet the DNS challenge's real-time requirement). 



In detail, given the magnitude spectrogram sequence $X$, network $G_{\mathrm{full}}$ firstly captures the global spectral pattern and the long-distance cross-band dependencies, and extracts a spectral embedding $\Tilde{x}_t$ for each frame. Such an embedding, is further concatenated with a contextual sub-band feature with center frequency bin of $f$. The concatenated representation $z_{t,f}$, together with its temporal dynamic feature, are then fed into network $G_{\mathrm{sub}}$ to learn local subband patterns. For each TF bin, $G_{\mathrm{sub}}$ yields a delayed cIRM \cite{Williamson16} $y_{t-\tau,f}\in \mathcal{R}^2$; lastly, by deploying $G_{\mathrm{sub}}$ on each frequency independently and sequentially, the cIRM of full spectrum is predicted for each frame. The whole process can be mathematically described below:
\begin{align*}
    X&=(x_1,\dots,x_t,\dots,x_T) \\
    \Tilde{x}_t &= G_{full}(X, x_t,\theta_{full})\\
    z_{t,f} &= [x_{t,f-N},\dots,x_{t,f},\dots,x_{t,f+N},\Tilde{x}_t] \\
    y_{t-\tau,f} &= G_{sub}(z_f, z_{t,f}, \theta_{sub})
\end{align*}

As reported in \cite{Hao21}, although both $G_{\mathrm{sub}}$ and $G_{\mathrm{full}}$ can achieve competitive performance, integration of them brings dramatically improved performance that exceeds top-ranked results in the DNS challenge.

\subsection{FullSubNet+}
As an extension of FullSubNet, FullSubNet+ \cite{fullsubnet+} has been proposed very recently. It claims two potential issues presented in the FullSubNet: a) feeding original whole frequency band to $G_{\mathrm{full}}$ might degrade network's discriminativity; and b) mapping from magnitude information to cIRM has the input-output mismatch issue. 

To address these issues, the following modifications are proposed: 1) extending the network with additional input of real and imaginary spectrograms; 2) introducing three parallel MulCA (multi-scale channel attention) modules before $G_{\mathrm{full}}$; 3) deploying two additional $G_{\mathrm{full}}$ modules to extract global information from real and imaginary domain, respectively; and 4) replacing LSTM with temporal convolutional network blocks in each $G_{\mathrm{full}}$ module to reduce model size. As a result, the proposed FullSubNet+ outperforms both FullSubNet and other speech enhancement methods.

%

\section{Proposed System}
Motivated by the success of those subband-based SE networks, we attempt to design a new subband-based SE system, PT-FSE. Based on the original architecture of FullSubNet, three new schemes are proposed herein to enhance the baseline.

\begin{figure*}[t]
  \centering
  \includegraphics[scale=0.58]{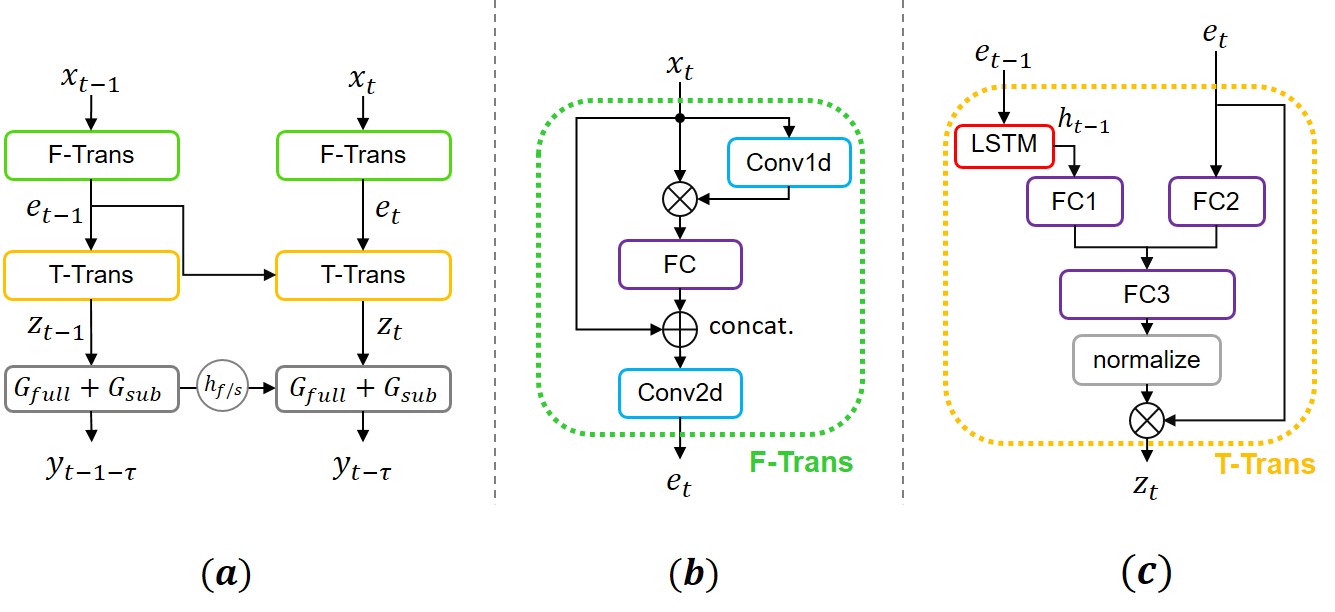}
  \caption{Structures of the proposed system: (a) overview of this speech enhancement system; (b) the Frequency Transformation Module (F-Trans); (c) the Temporal Transformation Module (T-Trans)}
  \label{fig:overview}
\end{figure*}

The network overview of PT-FSE is shown in fig. \ref{fig:overview} (a). Comparing to the FullSubNet, there are two additional modules, frequency transformation and temporal transformation, attached before the $G_{\mathrm{full}}$ network. 
In addition to these two additional transformation modules, a novel SE loss scheme is proposed to encourage the network to pay more attention for low-frequency SE.

\subsection{Frequency Transformation (F-Trans)}
Noting that the network $G_{\mathrm{full}}$ extracts the global representation from the raw full-band spectrum by LSTM and linear layers, such a design might degrade network’s discriminative ability among different frequency bands in the input spectrogram. This observation shares the motivation behind FullSubNet+ \cite{fullsubnet+}. However, different from exploiting the MulCA technique, herein we propose a scheme to insert the additional F-Trans module prior to the network $G_{\mathrm{full}}$. 

In fact, similar idea has been explored before in prior arts. In CRNN \cite{tan18}, a convolutional layer prior to the recurrent layer is adopted; and in PHASEN \cite{Yin20}, frequency transformation blocks (FTB) are proposed to capture long-range correlations along the frequency axis. 

Inspired by the FTB, the F-Trans module aims to transform the raw magnitude spectrum to yield a spectrum with better global frequency correlation. As illustrated in fig. \ref{fig:overview}(b), the raw magnitude spectrum at each frame is firstly processed by a single 1-dimensional convolution layer to generate frequency attention weights. The attention weighted spectrum, by focusing more on salient regions, is then sent into a fully connected layer; its output, concatenated with the raw spectrum, is finally pumped into a 2-dimensional convolution layer to generate frequency transformed magnitude spectrum.

\subsection{Temporal Transform (T-Trans)}
Apart from the additional frequency transformation, we propose to apply a temporal transform (T-Trans) module as well to benefit from incorporating long range temporal contexts. The motivation behind is the intuition that not all frames contribute equally to the enhancement and as proven in many prior SE studies, extracting contextual information over consecutive time frames is  beneficial for SE.  

Inspired by the work in \cite{SongLXZL16}, where a spatial-attention structure is proposed to emphasis discriminative changing joints. The idea is extended herein aiming to capture the amplitude evolution along temporal axis. As shown in fig. \ref{fig:overview}(c), the T-Trans takes the output from the F-Trans as input, and outputs temporal updated spectrum, with the same size of raw spectrum. Specifically, the previous input $e_{t-1}$, is sent into an LSTM layer. The last hidden state of the LSTM $h_{t-1}$, including the contextual information, is further mapped by a fully-connected layer. The projected output, together with another projected output of $e_t$, are merged and sent to the 3rd fully-connect layer. Followed by an normalization operation, the resulting normalized output is employed as the weights to update current input $e_t$. The weighted output serves as the input to the network $G_{\mathrm{full}}$.

Regarding the order of the proposed two modules, from our preliminary experiments, we find that placing the T-Trans module after the F-Trans achieves the best performance.

\subsection{PP-cIRM loss}

As introduced before, the baseline system adopts the $L_2$ loss (aka MSE loss)  of all cIRM at each TF bin as the learning target. Here the cIRM \cite{Williamson16} is defined as:
\begin{equation*}
M = \frac{Y_rS_r+Y_iS_i}{Y_r^2+Y_i^2} + i \frac{Y_rS_i-Y_iS_r}{Y_r^2+Y_i^2}   
\end{equation*}
where $S_r,S_i,Y_r,Y_i\in \mathcal{R}$ represent the real and imaginary component of complex spectrum of clean and noisy speech, respectively.  
By simultaneously enhancing both magnitude and phase spectra in the complex domain, the cIRM is an effective target for SE and has been widely adopted in numerous SE studies. 

Despite its popularity, in our opinion, cIRMs from each TF bin share the identical contribution weight to the overall loss, which does not match human perception. In fact, it is well known that human ear is most sensitive in the range of about 1–5 kHz, and get decreased sensitivity at higher and lower frequencies \cite{book07}. As such, simple $L_2$ loss of cIRMs is not the optimal objective to optimize and may lead to wasted model capacity and limited performance. 

To address the problem, variant perceptually motivated losses have been proposed in recent SE works. For example, PercepNet \cite{2020arXiv200804259V} is proposed by focusing on the spectral envelope and the periodicity of the speech; to incorporate a speech perception model into the loss function, a short-time objective intelligibility metric is introduced \cite{zhaoyan2018}, to name a few.

In contrast to these methods, we propose to replace the original MSE loss with a new perceptual loss, with core motive to leverage the knowledge of human hearing perception.

\begin{figure}[t]
  \centering
  \includegraphics[scale = 0.38]{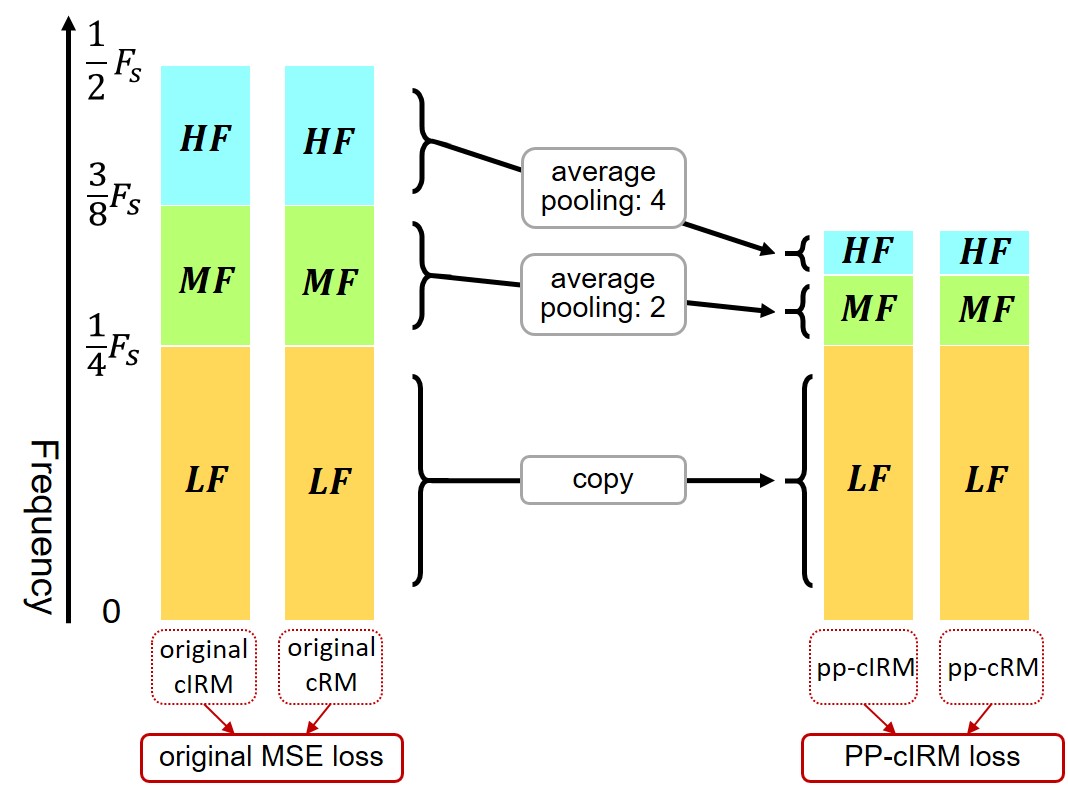}
  \caption{Illustration of PP-cIRM loss}
  \label{fig:pp-loss}
\end{figure}

Motivated by the concept of critical bandwidth, where the bandwidth of the filter increases with the frequency, we come up with a simple implementation of proposed perceptual loss, called Pyramid Pooling based cIRM (PP-cIRM). It is designed to calculate the MSE at decreasingly granularity scale from high-frequency to low-frequency regions. The detailed loss construction is illuminated in fig. \ref{fig:pp-loss}. Specifically, given a full-band spectrum with $F_s/2$ frequency bins, it is firstly delineated into 3 patches: LF (low-frequency: $0 \sim \mathrm{F_s}/4$), MF (middle-frequency: $\mathrm{Fs}/4 \sim 3\mathrm{F_s}/8$) and HF (high-frequency: $3\mathrm{F_s}/8 \sim \mathrm{F_s}/2$); then cIRMs at both MF and HF are pooled with scale of 2 and 4 respectively; finally, the MSE loss is calculated based on pooled cIRMs. 

The whole construction process of the new PP-cIRM can be mathematically formulated as:
\begin{equation}
 \begin{aligned}
  L_{_\mathrm{PP-cIRM}} & = \ w_1*{\left\| \mathrm{cIRM}_{_\mathrm{LF}} - \mathrm{cRM}_{_\mathrm{LF}} \right\|}_{2} \\
            &+ w_2*{\left\| f_{{pool2}}(\mathrm{cIRM}_{_\mathrm{MF}}) - f_{{pool2}}(\mathrm{cRM}_{_\mathrm{MF}}) \right\|}_{2} \\ 
            &+ w_3*{\left\| f_{{pool4}}(\mathrm{cIRM}_{_\mathrm{HF}}) - f_{{pool4}}(\mathrm{cRM}_{_\mathrm{HF}}) \right\|}_{2}
  \label{eq1}
 \end{aligned}
\end{equation}
where $f_{pool2}()$ denotes the average pooling function with step size 2 and $f_{pool4}()$ with step size 4. 

It is worth noting that the core idea of PP-cIRM is to guide the model to focus more on TF bins with finer scale (i.e., low frequency bins). We understand that different PP-cIRM can be implemented with sophisticated pyramid scales or pooling strategies. In this study, for simplicity, only the loss formulated above is implemented and evaluated in our experiments.


\begin{table*}[tb]
  \renewcommand\arraystretch{1.36}
  \caption{Performance comparisons for different SE systems on the DNS Challenge test dataset}
  \label{tab:dns2020_1}
  \centering
  \begin{tabular}{l |cccc | cccc}
    
    \hline
    \hline
    
    \multirow{2}{*}{Methods}
        &\multicolumn{4}{c|}{with-reverb}  &\multicolumn{4}{c}{without-reverb} \\
       \cline{2-5} \cline{6-9} 
        &WB-PESQ      &NB-PESQ      &STOI      &SI-SDR    &WB-PESQ      &NB-PESQ      &STOI      &SI-SDR \\
    \hline

    Noisy    &1.822     &2.753       &86.62       &9.033    & 1.582     &2.454       &91.52       &9.071 \\
    PoCoNet\cite{poconet20} \footnotesize{(2020)}  & 2.832  & -  &    -   &    -    &    2.748       & -     &  -  &  -   \\
    DCCRN\cite{hu20g} \footnotesize{(2021)} &  -  & 3.20  &    -   &    -    &    -        & {3.26}     &  -  &  -   \\
    DCCRN+\cite{Lv21} \footnotesize{(2021)} &  -  & 3.30  &    -   &    -    &    -        & {3.330}     &  -  &  -   \\
    HGCN\cite{wang22} \footnotesize{(2022)}    &    -    &   -    &    -    &   -    & {2.883}      &   -   & {96.50}       &{18.144}      \\
    
    \hline
    FullSubNet\cite{Hao21} \footnotesize{(2021)}  & {2.969}    & {3.473}      & {92.62}      & {15.750}  & 2.777      & 3.305    &96.11       &17.290      \\
    Baseline (our implemented FullSubNet) &3.007 &3.512 &92.85 &16.04 &2.845 &3.347 & 96.16 & 17.33 \\
    FullSubNet+\cite{fullsubnet+} \footnotesize{(2022)}     & \underline{3.218}    & \textbf{3.666}      & \textbf{93.84}      & \underline{16.81}     & \underline{2.982}      & \underline{3.504}     & \underline{96.69}       & \underline{18.340}      \\
    OURS     & \textbf{3.219}    & \underline{3.620}      & \underline{93.82}      & \textbf{16.87}     & \textbf{3.060}      & \textbf{3.512}     & \textbf{96.90}       &\textbf{18.432}      \\

    \hline
    \hline
  \end{tabular}
\end{table*}

\section{Experiments}

\subsection{Dataset and System Setting}

\textbf{Dataset} The DNS challenge dataset \cite{Reddy20}, as a popular benchmark dataset to evaluate SE systems, is used for our experiments. For fair comparison with our baseline system, we following the same configuration of training and evaluation dataset and the data preparation strategy as in \cite{Hao21}. 

The training data in the dataset contains over 500 hours of clean English speech across 2150 speakers and up to 180 hours noise clips from 150 classes; and the same dynamically mixing strategy is adopted for the training preparation. As for the evaluation, the test set is categorised into two subsets, under the condition of with and without reverberation, respectively; and each test subset contains 150 clips with SNR ranging from 0dB to 20dB. 

\noindent\textbf{Implementation details}
Firstly, all speech signals are re-sampled to sampling rate of 16kHz. Window length of 512 samples is used for FFT with shift size of 256 samples. Two-frame look ahead (i.e., $\tau=2$ in fig. \ref{fig:FSN} is allows, which corresponds to 32 ms and meets the official requirement in real-time DNS. 
Our proposed PT-FSE system is implemented with the following details:
\begin{itemize}
\item FPM module: the two conv2d layers have different channel sizes (1,1) and (2,1) and share same kernel size of (1,1); and the conv1d layer has kernel size of 9;
\item TPM module: three FC layers all have 257 neurons; the LSTM layer has hidden size of 257;
\item FullSubNet module: with identical network configuration as the original baseline, the network $G_{\mathrm{full}}$ has 2-layer LSTM with hidden size of 512 and $G_{\mathrm{full}}$ has the same structure with size of 384.
\end{itemize}

For system training, we use pytorch and choose Adam as the optimizer. The learning rate is set to $0.001$ for the first 100 epochs and then decreased to $0.0003$ afterwards. PP-cIRM weights are set to $\{w_1,w_2,w_3\}=\{1,1,1\}$ by default without further fine-tuning. With batch size of 42, the drop band strategy is also employed.

\subsection{Evaluation}
We follow the same evaluation method and metrics as in most prior arts. That is, our system is evaluated with metrics of PESQ, STOI, and SI-SDR. Note that in DNS challenge dataset, PESQ can be measured on basis of either narrow band (NB-PESQ) or wide band (WB-PESQ). In this study, we report both metrics in order to make sufficient performance comparisons with more prior arts. 

Firstly, our baseline system is constructed by re-producing the codes provided by authors\footnote{\url{https://github.com/haoxiangsnr/FullSubNet}}. The re-produced baseline system is on par with the original FullSubNet, with slight performance improvement. Then our system is built by implementing all proposed schemes On top of the baseline. 
 
Tab.\ref{tab:dns2020_1} lists the performance comparison results of our system with prior arts on the DNS challenge test set. For highlight purpose, the best performance is bold-faced and the 2nd best is underlined. The vacant place indicates that the corresponding result is not reported in the original paper. 

Note that all prior arts listed in Tab.\ref{tab:dns2020_1} have excellent performances. The PoCoNet and DCCRN ranked top two in 2020 DNS Challenge and DCCRN+ is an advanced version of DCCRN and outperforms DCCRN. HGCN\cite{wang22} is a very successful method published very recently that surpasses the DCCRN+. In addition, to ensure the fairness, performance scores are directly quoted from the original papers. 

A quick glance at the results shows a clear trend that subband-based SE systems (lower part) perform better than competitive prior arts (upper part), which confirms our early statement that to enhance each subband individually is appealing and effective. In addition, focusing on the subband-based systems, we have the following observations:
\begin{itemize}
    \item on the with-reverb dataset:
    \begin{itemize}
        \item our proposed system significantly outperforms its baseline, with relative improvement of $7\%$ and $3.7\%$ regarding the PESQ scores;
        \item our proposed system is on par with previous best SOTA, FullSubNet+ system, being neck and neck on performance scores across all objective metrics;
    \end{itemize}  
    \item on the without-reverb dataset:
    \begin{itemize}
         \item our proposed system shows more performance gains over its baseline (relative improvement of $9.5\%$ and $5.4\%$ on PESQ), comparing to the case of with-reverb;
        \item FullSubNet+ system is the previous best SOTA;
        \item our proposed system outperforms the FullSubNet+ over all metrics;
    \end{itemize}
\end{itemize}

In all, by considering both with- and without-reverb cases, our proposed system yields the best performance on DNS 2020 test dataset, with average WB-PESQ of 3.140 and average NB-PESQ of 3.566. It is worth noting that such performance is achieved under the condition of a smaller model size, comparing to the previous SOTA system, FullSubNet+ (as explained in next subsection). 

\subsection{Ablation study}
A series of ablation studies is conducted to reveal individual contribution of our proposed three schemes. Our experiments show that experimental results on with- and without-reverb share the similar trend. As such, for simplicity, only the results on the without-reverb scenario are reported.

There are three proposed schemes, so our baseline system is updated by taking one component at a time. All the resulting systems are measured by the same metrics of WB-PESQ, NB-PESQ, STOI and SI-SDR, with performance reported in Tab.\ref{tab:ablation}. Here +\emph{F-Trans}, +\emph{T-Trans}, +\emph{F-Trans}+\emph{T-Trans} and +\emph{PP-cIRM} denote the updated baseline systems by incorporating the proposed \emph{F-Trans}, \emph{T-Trans}, \emph{F-Trans}+\emph{T-Trans} and \emph{PP-cIRM} components, respectively; and \emph{FULL} represents the our proposed system.

\begin{table}[htb]
  \renewcommand{\arraystretch}{1.35}
  \setlength\tabcolsep{4.5pt}
  \caption{Ablation study results using the test dataset without reverberation}
  \label{tab:ablation}
  \centering

  \begin{tabular}{l | ccccc}
    
    \hline
    \hline
    
    \multirow{2}{*}{Model}      &\multirow{2}{*}{Size(M)}     &\multirow{2}{*}{WB-PESQ}      &\multirow{2}{*}{NB-PESQ}        &\multirow{2}{*}{STOI}        &\multirow{2}{*}{SI-SDR}       \\
    \\
    
    \hline
    
    Baseline    & 5.64   & 2.845      & 3.347    &96.16       &17.33     \\
    +\emph{F-Trans}      & 5.84    & 2.921      & 3.403    &96.54       &17.60      \\
    +\emph{T-Trans}      & 6.14    & 2.908      & 3.392    &96.44       &17.63      \\
    +\emph{PP-cIRM}  & 5.64    & 2.848      & 3.358    &96.20       &17.49      \\
    +\emph{F-Trans}+\emph{T-Trans}  & 6.34    & 2.933      & 3.410     & 96.55       &17.54      \\
    FULL        & 6.34   & \textbf{3.060}      & \textbf{3.512}     & \textbf{96.90}       &\textbf{18.43}      \\
    \hline
    \hline
  \end{tabular}
\end{table}

\begin{figure}[htb]
  \centering
  \includegraphics[scale=0.30]{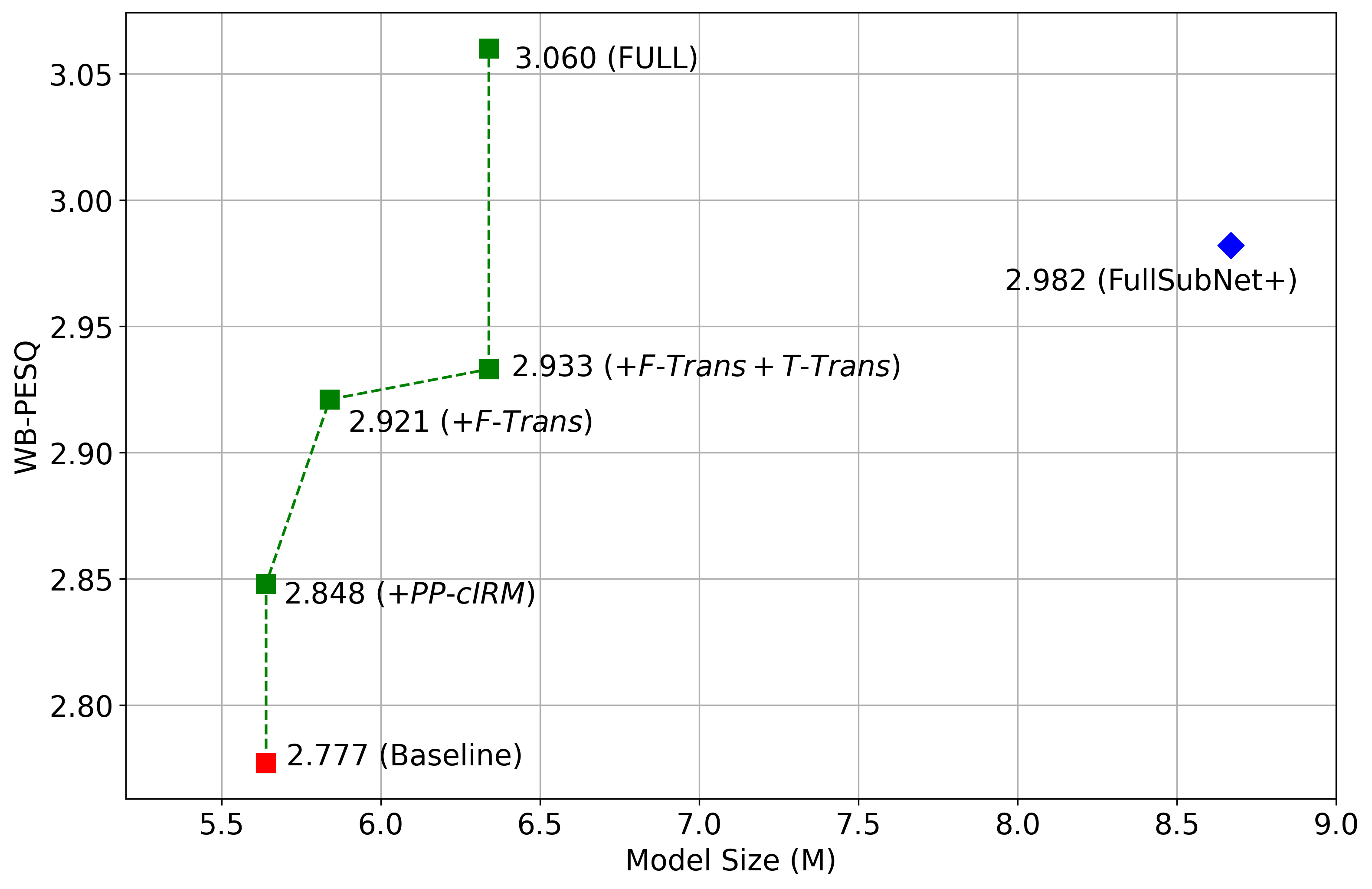}
  \caption{Relation between system performance and capacity }
  \label{fig:ablation}
\end{figure}

\begin{figure}[htb]
  \centering
  \includegraphics[scale=0.40]{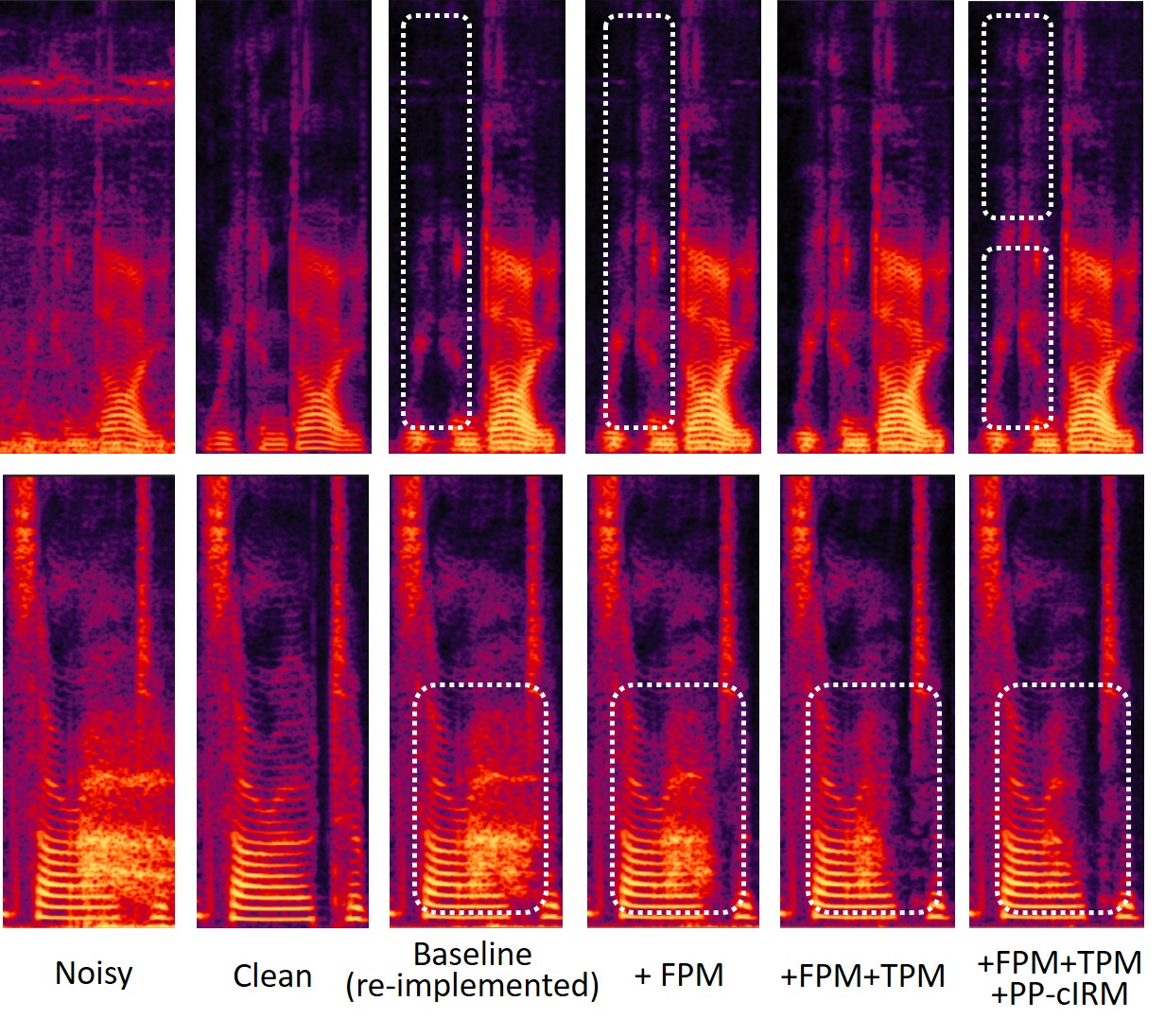}
  \caption{Effects comparison of the proposed schemes via spectrogram Visualization}
  \label{fig:spec}
\end{figure}

As we can observe, choosing WB-PESQ as a representative, the proposed F-Trans, T-Trans and PP-cIRM scheme all boost performance of the baseline system, with respective score contribution of $2.67\%$, $2.22\%$, $0.11\%$ relatively. It is interesting to note that the TF-Transformed baseline, by adding combined  \emph{F-Trans} and \emph{T-Trans} on the baseline, achieves a significant performance gain ($3.09\%$). This shows that the two transformation are complementary for enhancement. Furthermore, by adding PP-CIRM on the TF-Transformed baseline, a larger margin ($4.46\%$) is obtained, which again, confirms that all three proposed schemes are complementary to each other.


Apart from the performance, we also examine the relationship between model performance and capacity. Such a relationship is depicted in Fig. \ref{fig:ablation} as one scatter diagram, where x-axis shows the model size and y-axis denotes the WB-PESQ score. The official FullSubNet system (in red square) shows excellent performance with the smallest model size of 5.64M. Using it as an anchor system, our proposed system (in green squares) achieves the best performance with reasonable size increment of 0.7M (0.5M from T-Trans and 0.2M from F-Trans); and the FullSubNet+ (in blue diamond) achieves intermediate performance with the cost of high capacity overhead (up to 8.67M) .




Finally, to visualize the impact of our proposed schemes, spectrograms of two noisy examples, randomly sampled from test speech clips, are presented in fig. \ref{fig:spec}. For both clips, from left to right as more proposed schemes are applied, we can visually observe that the extracted speech is progressively enhanced with reference to the target (i.e., spectrogram of clean speech). Specifically speaking, regarding the speech in the upper panel, the missed information (highlighted by the dashed white box) presented in our baseline is restored progressively; and for the speech in the lower panel, the additive noise and smearing effect (in the dashed white box) presented in our baseline is removed progressively. These observations, from another angle, demonstrate that our schemes are effective and can complement each other. More enhanced audio examples can be accessed online\footnote{\url{https://xuchengone.github.io/PT-FSE-github.io/}}.


\section{Conclusions}
In this paper, we present a subband-based monaural speech enhancement system, termed PT-FSE. It adopts the FullSubNet, a representative work of subban-based SE methods, as the baseline. On top of it, we propose PT-FSE with three updates to further strengthen the strong baseline. As proven by experimental results on the DNS2020 dataset, the proposed PT-FSE system significantly outperforms its baseline system across various metrics, including PESQ and STOI measures. Besides findings on the model effectiveness, our ablation studies also confirm that the proposed three updates are mutually beneficial for the system performance. Furthermore, we expand the scope of our benchmark system to include a few systems with top rankings on the DNS2020 dataset. Our system consistently surpasses these competitive systems and provides the best perceptual result reported till date, in terms of average NB-PESQ. Looking forward, we are interested in extending the proposed system to favor the downstream automatic speech recognition task as well, an interesting yet challenging topic for our future investigations. 


\begin{thebibliography}{10}
\providecommand{\url}[1]{#1}
\csname url@samestyle\endcsname
\providecommand{\newblock}{\relax}
\providecommand{\bibinfo}[2]{#2}
\providecommand{\BIBentrySTDinterwordspacing}{\spaceskip=0pt\relax}
\providecommand{\BIBentryALTinterwordstretchfactor}{4}
\providecommand{\BIBentryALTinterwordspacing}{\spaceskip=\fontdimen2\font plus
\BIBentryALTinterwordstretchfactor\fontdimen3\font minus
  \fontdimen4\font\relax}
\providecommand{\BIBforeignlanguage}[2]{{%
\expandafter\ifx\csname l@#1\endcsname\relax
\typeout{** WARNING: IEEEtran.bst: No hyphenation pattern has been}%
\typeout{** loaded for the language `#1'. Using the pattern for}%
\typeout{** the default language instead.}%
\else
\language=\csname l@#1\endcsname
\fi
#2}}
\providecommand{\BIBdecl}{\relax}
\BIBdecl

\bibitem{Luo2017}
Y.~Luo and N.~Mesgarani, ``Tasnet: Time-domain audio separation network for
  real-time, single-channel speech separation,'' in \emph{2018 IEEE
  International Conference on Acoustics, Speech and Signal Processing
  (ICASSP)}, 2018, pp. 696--700.

\bibitem{zhang20m_interspeech}
Z.~Zhang, B.~He, and Z.~Zhang, ``{X-TaSNet: Robust and Accurate Time-Domain
  Speaker Extraction Network},'' in \emph{Proc. Interspeech 2020}, 2020, pp.
  1421--1425.

\bibitem{Luo2019}
Y.~Luo and N.~Mesgarani, ``Conv-tasnet: Surpassing ideal time–frequency
  magnitude masking for speech separation,'' \emph{IEEE/ACM Transactions on
  Audio, Speech, and Language Processing}, vol.~27, p. 1256–1266, Aug 2019.

\bibitem{Luo20}
Y.~Luo, Z.~Chen, and T.~Yoshioka, ``Dual-path rnn: Efficient long sequence
  modeling for time-domain single-channel speech separation,'' \emph{ICASSP
  2020 - 2020 IEEE International Conference on Acoustics, Speech and Signal
  Processing (ICASSP)}, pp. 46--50, 2020.

\bibitem{zhang21}
Z.~Zhang, B.~He, and Z.~Zhang, ``Transmask: A compact and fast speech
  separation model based on transformer,'' \emph{ICASSP 2021 - 2021 IEEE
  International Conference on Acoustics, Speech and Signal Processing
  (ICASSP)}, pp. 5764--5768, 2021.

\bibitem{Hummersone14}
C.~Hummersone, T.~Stokes, and T.~Brookes, ``On the ideal ratio mask as the goal
  of computational auditory scene analysis,'' in \emph{Blind Source Separation:
  Advances in Theory, Algorithms and Applications}, 05 2014, pp. 349--368.

\bibitem{tan18}
K.~Tan and D.~Wang, ``{A Convolutional Recurrent Neural Network for Real-Time
  Speech Enhancement},'' in \emph{Proc. Interspeech 2018}, 2018, pp.
  3229--3233.

\bibitem{choi19}
H.-S. Choi, J.-H. Kim, J.~Huh, A.~Kim, J.-W. Ha, and K.~Lee, ``Phase-aware
  speech enhancement with deep complex u-net,'' \emph{ArXiv}, vol.
  abs/1903.03107, 2019.

\bibitem{Tan19}
K.~Tan and D.~Wang, ``Complex spectral mapping with a convolutional recurrent
  network for monaural speech enhancement,'' in \emph{2019 IEEE International
  Conference on Acoustics, Speech and Signal Processing (ICASSP)}, 2019, pp.
  6865--6869.

\bibitem{hu20g}
Y.~Hu, Y.~Liu, S.~Lv, M.~Xing, S.~Zhang, Y.~Fu, J.~Wu, B.~Zhang, and L.~Xie,
  ``{DCCRN: Deep Complex Convolution Recurrent Network for Phase-Aware Speech
  Enhancement},'' in \emph{Proc. Interspeech 2020}, 2020, pp. 2472--2476.

\bibitem{Lv21}
S.~Lv, Y.~Hu, S.~Zhang, and L.~Xie, ``{DCCRN+: Channel-Wise Subband DCCRN with
  SNR Estimation for Speech Enhancement},'' in \emph{Proc. Interspeech 2021},
  08 2021, pp. 2816--2820.

\bibitem{Hao21}
X.~Hao, X.~Su, R.~Horaud, and X.~li, ``Fullsubnet: A full-band and sub-band
  fusion model for real-time single-channel speech enhancement,'' in
  \emph{ICASSP 2021}, 2021, pp. 6633--6637.

\bibitem{Li19}
X.~Li and R.~Horaud, ``Narrow-band deep filtering for multichannel speech
  enhancement,'' \emph{ArXiv}, vol. abs/1911.10791, 2019.

\bibitem{Yin20}
D.~Yin, C.~Luo, Z.~Xiong, and W.~Zeng, ``{PHASEN}: A phase-and-harmonics-aware
  speech enhancement network,'' \emph{Proceedings of the AAAI Conference on
  Artificial Intelligence}, vol.~34, pp. 9458--9465, 2020.

\bibitem{dns2020}
C.~Reddy, V.~Gopal, R.~Cutler, E.~Beyrami, R.~Cheng, H.~Dubey, S.~Matusevych,
  R.~Aichner, A.~Aazami, S.~Braun, P.~Rana, S.~Srinivasan, and J.~Gehrke, ``The
  interspeech 2020 deep noise suppression challenge: Datasets, subjective
  testing framework, and challenge results,'' in \emph{Proc. Interspeech 2020},
  05 2020.

\bibitem{fullsubnet+}
\BIBentryALTinterwordspacing
J.~Chen, Z.~Wang, D.~Tuo, Z.~Wu, S.~Kang, and H.~Meng, ``Fullsubnet+: Channel
  attention fullsubnet with complex spectrograms for speech enhancement,''
  \emph{CoRR}, vol. abs/2203.12188, 2022. [Online]. Available:
  \url{https://doi.org/10.48550/arXiv.2203.12188}
\BIBentrySTDinterwordspacing

\bibitem{Williamson16}
D.~S. Williamson, Y.~Wang, and D.~Wang, ``Complex ratio masking for monaural
  speech separation,'' \emph{IEEE/ACM Transactions on Audio, Speech, and
  Language Processing}, vol.~24, no.~3, pp. 483--492, 2016.

\bibitem{SongLXZL16}
S.~Song, C.~Lan, J.~Xing, W.~Zeng, and J.~Liu, ``An end-to-end spatio-temporal
  attention model for human action recognition from skeleton data,'' in
  \emph{Proceedings of the Thirty-First AAAI Conference on Artificial
  Intelligence}, ser. AAAI'17.\hskip 1em plus 0.5em minus 0.4em\relax AAAI
  Press, 2017, p. 4263–4270.

\bibitem{book07}
\BIBentryALTinterwordspacing
P.~C. Loizou, \emph{Speech Enhancement: Theory and Practice}.\hskip 1em plus
  0.5em minus 0.4em\relax CRC Press, 2013. [Online]. Available:
  \url{https://doi.org/10.1201/b14529}
\BIBentrySTDinterwordspacing

\bibitem{2020arXiv200804259V}
J.-M. {Valin}, U.~{Isik}, N.~{Phansalkar}, R.~{Giri}, K.~{Helwani}, and
  A.~{Krishnaswamy}, ``{A Perceptually-Motivated Approach for Low-Complexity,
  Real-Time Enhancement of Fullband Speech},'' \emph{arXiv e-prints}, p.
  arXiv:2008.04259, Aug. 2020.

\bibitem{zhaoyan2018}
Y.~Zhao, B.~Xu, R.~Giri, and T.~Zhang, ``Perceptually guided speech enhancement
  using deep neural networks,'' in \emph{2018 IEEE International Conference on
  Acoustics, Speech and Signal Processing (ICASSP)}, 2018, pp. 5074--5078.

\bibitem{poconet20}
U.~Isik, R.~Giri, N.~Phansalkar, J.-M. Valin, K.~Helwani, and A.~Krishnaswamy,
  ``Poconet: Better speech enhancement with frequency-positional embeddings,
  semi-supervised conversational data, and biased loss,'' in \emph{Proc.
  Interspeech 2020}, 2020, p. 2487–2491.

\bibitem{wang22}
\BIBentryALTinterwordspacing
T.~Wang, W.~Zhu, Y.~Gao, J.~Feng, and S.~Zhang, ``{HGCN: harmonic gated
  compensation network for speech enhancement},'' 2022. [Online]. Available:
  \url{http://arxiv.org/abs/2201.12755}
\BIBentrySTDinterwordspacing

\bibitem{Reddy20}
\BIBentryALTinterwordspacing
C.~K.~A. Reddy, E.~Beyrami, H.~Dubey, V.~Gopal, R.~Cheng, R.~Cutler,
  S.~Matusevych, R.~Aichner, A.~Aazami, S.~Braun, P.~Rana, S.~Srinivasan, and
  J.~Gehrke, ``The {INTERSPEECH} 2020 deep noise suppression challenge:
  Datasets,subjective speech quality and testing framework,'' \emph{Proc.
  Interspeech 2020}, 2020. [Online]. Available:
  \url{http://arxiv.org/abs/2001.08662}
\BIBentrySTDinterwordspacing

\end{thebibliography}

\end{document}